# Spectrum of the relic neutrino background from past supernovae and cosmological models

Tomonori Totani and Katsuhiko Sato

Department of Physics, Faculty of Science, the University of Tokyo
Bunkyo-ku, Tokyo 113, JAPAN

**Abstract**

It is greatly expected that the relic neutrino background from past supernovae is detected by Superkamiokande (SK) which is now under construction. We calculate the spectrum and the event rate at SK systematically by using the results of simulations of a supernova explosion and reasonable supernova rates.

We also investigate the effect of a cosmological constant, $\Lambda$, on the spectrum, since some recent cosmological observations strongly suggest the existence of $\Lambda$.

We find following results. 1) The spectrum has a peak at about 3 MeV, which is much lower than that of previous estimate (6 ∼ 10 MeV). 2) The event rate at SK in the range from 10 MeV to 50 MeV, where the relic neutrinos from past supernovae is dominant, is about $25\, h_{50}{}^2 \left(\frac{R_{SN}}{0.1\,\mathrm{yr}^{-1}}\right) \left(\frac{n_G\, h_{50}^{-3}}{0.02\mathrm{Mpc}^{-3}}\right)$ events per year, where $R_{SN}$ is the supernova rate in a galaxy, $n_G$ is the number density of galaxies, and $h_{50} = H_0/ (50\mathrm{km/s/Mpc})$, where $H_0$ is the Hubble constant. 3) The event rate is almost insensitive to $\Lambda$. The flux increases in the low energy side ($<$ 10 MeV) with increasing $\Lambda$, but decreases in the high energy side (10 MeV $<$) in models in which the integrated number of supernovae in one galaxy is fixed.

# 1  Introduction

It is generally believed that the type II supernova explosions have occurred everywhere in the universe since galaxies began to form, and a great number of neutrinos emitted from such supernovae should make a diffuse background. This *relic neutrino background from past supernovae* has been discussed by some researchers (Bisnovatyi-Kogan and Seidov [1], Krauss, Glashow, and Schramm [2], Woosley, Wilson, and Mayle [3], Hirata [4], Totsuka [5], and Suzuki [6]), however, only rough estimates of the flux or event rates at specific detectors are given in previous papers. In the near future, an international network of deep underground neutrino telescopes will be constructed and the detectability will be remarkably improved. The water Čerenkov detector, Superkamiokande (SK, under construction) has particularly bright future among these projects and the relic neutrinos are considered to be one of the targets of this detector. Although the background has not yet been observed (upper limit $\leq 780\bar{\nu}_e$ /cm$^2$/s has obtained from the Kamiokande II experiment [4]), we can expect the detection in the next decade. Therefore it is quite important to predict the background flux and response of detectors as accurately as possible.

We calculate the spectrum of the relic neutrinos using the results of simulations of a supernova explosion for some values of stellar mass, and reasonable supernova rates. We also calculate the event rate at SK for the flux of all species of neutrinos as a function of energy of recoil electrons (or positrons).

The universe with non vanishing cosmological constant, $\Lambda$, has not been considered in any previous work [1–6]. However, recently the Hubble Space Telescope (Freedman et al. [7]) found a lot of Cepheids in a galaxy of the Virgo cluster M100, and derived the Hubble constant $80 \pm 17$ km/s/Mpc from the Cephieds observation. Pierce et al. [8] also reported the value $87 \pm 7$ km/s/Mpc from ground-based observations of Cephieds. Such uncomfortably large values suggest the age of the universe to be roughly 10 Gyr in the standard cosmology without a cosmological constant. But recent values based upon dating the ages of the globular clusters of old stars are rather high, for example 15 Gyr $\pm$ 3 Gyr [9]. The simplest solution to this problem is to include $\Lambda$. Existence of $\Lambda$ is also suggested from other cosmological observations. The dynamical methods strongly indicate that the material such as clusters or galaxys on scales less than about 10 to 30 Mpc contributes $\Omega_0 \simeq 0.2$ [10]. On the other hand, the inflationary cosmology requires $\Omega_0 + \lambda = 1$ ($\lambda = \frac{\Lambda}{3H_0^2}$, $H_0$ is the Hubble constant), then there must be non vanishing $\Lambda$. Furthermore, there is an observation which suggests the existence of a cosmological constant: the test with the number count of faint galaxies performed by Fukugita, Takahara, Yamashita, Yoshii [11].

Therefore it is not up-to-date to calculate the relic neutrino flux neglecting $\Lambda$, so we calculate the spectrum and the event rate with some values of $\lambda$ under $\Omega_0 + \lambda = 1$.

Section 2 is devoted to construction of a model for neutrino emission from a supernova and the supernova rate in a galaxy. The fluxes of the neutrino background with some values of $\Lambda$ are shown in section 3, and compared with fluxes of ohter sources of neutrinos. In section 4, We predict the response of the SK detector. Response of the heavy-water Čerenkov detector, Sudbury Neutrino Observatory (SNO) is also discussed briefly. We summarize our results in section 5.

## 2 Model of Neutrino Emission and Supernova Rate

### 2.1 Neutrino Emission from a Supernova

It is generally believed that each type of neutrinos carries away approximately the same amount of energy from thermal cooling phase. However, the flavor which is most easily detected in normal water Čerenkov detectors is $\bar{\nu}_e$ (discussed in more detail in section 4), so we consider mainly $\bar{\nu}_e$ throughout this paper.

We assume the energy distribution of emitted $\bar{\nu}_e$'s to be the Fermi-Dirac (F-D) distribution. We also assume that chemical potential, $\mu = 0$. Although there are some fittings which include $\mu$, such fittings generally give smaller temperature [12]. Therefore we think that the assumption is reasonable with the temperature used below. Then time-integrated amount of emitted neutrinos can be determined if temperature $T_\nu$ and total emitted energy of a supernova $E_\nu (= \int L_\nu dt$, $L_\nu$ is neutrino luminosity) are given.

The number of emitted $\bar{\nu}_e$'s of energy $q$ from a supernova per unit $q$, $\frac{dN_\nu(q)}{dq}$, can be obtained after easy calculation:

$$\frac{dN_\nu(q)}{dq} = \frac{c}{4} \frac{4\pi}{(2\pi\hbar c)^3} \frac{q^2}{e^{\beta(q-\mu)}+1} 4\pi r_\nu^2 \Delta t_{burst} \, , \tag{1}$$

where $r_\nu$ and $\Delta t_{burst}$ represent the radius of neutrinosphere and the duration of the burst, respectively. these two parameters should be determined from the values of $E_\nu, T_\nu$, which generally depend upon the mass of the progenitor.

We use the results of the numerical calculations made by Woosley, Wilson and Mayle [3]. According to them, these values for $\bar{\nu}_e$'s are $E_\nu = 4.8, 6.0, 11$ ($10^{52}$ergs) and $T_\nu = 4.0, 5.0, 5.3$ (MeV) for the stellar masses of $10 M_\odot, 15 M_\odot, 25 M_\odot$, respectively. We assume that all stars of masses above $8 M_\odot$ end their life with type II supernova explosions, and consider the mass range from $8 M_\odot$ to $50 M_\odot$. Each star in the mass range $8 M_\odot \sim 12.5 M_\odot, 12.5 M_\odot \sim 20 M_\odot, 20 M_\odot \sim 50 M_\odot$ is regarded as a star of a mass $10 M_\odot, 15 M_\odot, 25 M_\odot$, respectively. To get the weighted average value, we also have to determine the stellar population as a function of mass. We assume that the mass distribution of type II supernova is the same as the initial stellar mass function. The analytic fit given by Shapiro and Teukolsky [13] is adopted, which is a modern version of the Salpeter's mass function [14]:

$$\log \xi(\log M) = A_0 + A_1 \log M + A_2 (\log M)^2, \tag{2}$$

where $A_0 = 1.41, A_1 = -0.90, A_2 = -0.28, \log M > -1$. Here $\xi$ [pc$^{-2}$] is the total number of field stars that have *ever* formed, per logarithmic mass interval. Stellar mass $M$ is measured in units of $M_\odot$ and logarithms are to base 10. Note that almost all stars above $8M_\odot$ that have ever formed have already exploded.

Now we can calculate a weighted average energy distribution of the neutrinos emitted from a supernova. In Fig.1, the average distribution and the F-D distributions for the three values of the mass are shown. We use this weighted average distribution to calculate the flux of the $\bar{\nu}_e$ background.

## 2.2 Supernova Rate

While the supernova rate plays a crucial role in calculation of the flux of the neutrino background, the supernova rate and its time dependence is not well understood. There are some estimates of supernova rates based upon different methods. For example, about 1 per 28 years per a galaxy is obtained from the seven observed supernovae [15], 1 per 50~60 years from the radio pulsar data [16], 1 per 10 years [17] and 1 per 50 years [18] from statistics on stellar types. Furthermore, there are some estimates based upon nucleosynthesis arguments, which can avoid the statistical ambiguities. According to Arnett et al. [19], the rate is estimated to be about 1 per 10 years for our Galaxy. Woosley et al. [3] estimated the integrated number density of gravitational collapses to be $3.6 \times 10^{-67}(h_{50})^2 \left(\frac{\Omega_b}{0.1}\right)$ [cm$^{-3}$]. Here, $h_{50} = H_0/(50\text{km/s/Mpc})$. If we take the number density of galaxies as $0.02(h_{50})^3$[Mpc$^{-3}$] and the age of galaxies as $10^{10}$ years, this value results in the rate $1/18\ h_{50}^{-1}\text{yr}^{-1}$ per a galaxy.

Throughout this paper, we assume that the time integrated number of type II supernovae in a galaxy is $10^9$ according to the nucleosynthesis argument of Arnett et al. [19] (they derived the above rate 0.1 yr$^{-1}$ from this estimate and assuming the age of our galaxy to be $10^{10}$ yr). This value is based upon the abundance of oxygen which is produced chiefly in type II supernovae. The observational data for abundance of oxygen in the HII regions are compiled in Díaz [20]. As for time dependence of the supernova rate, we consider following two simple models. In the first model (we call this 'constant model'), we assume that gravitational collapses began at the redshift parameter $z_{max} = 5$ and after that the rate is constant with time. Therefore, the supernova rate is $(10^9)/(\text{time from } z = z_{max} \text{ to } z = 0)$. In the second model (we call this 'constant+burst model'), we divide the rate into two components, i.e. the constant component and the burst component. The rate for each component is set so that the time integrated number of each component is $0.5 \times 10^9$ per a galaxy. The burst component is a gaussian about $z$ which is centered at $z = 2.25$ and its standard deviation $\Delta z$ is 0.25. These values are based upon the numerical calculations of a formation of a galaxy performed by Katz [21]. The supernova rates obtained in this way for the two models are shown in Table.1 with some values of $\lambda$, including the cosmic time from $z = 5$ to $z = 0$, the rates of the constant component of the two models, and the rates of the burst component at the burst peak.

## 3 Flux Calculations

### 3.1 Formulation

In order to get the differential number flux of background neutrinos, first we calculate the present number density of the background neutrinos per unit neutrino energy, $\frac{dn_\nu(q)}{dq}$, where $q$ is neutrino energy. The contribution of the neutrinos emitted in the interval of cosmic time $t \sim t + dt$ to $dn_\nu(q)$ is given by

$$dn_\nu(q) = n_G(t) R_{SN}(t) dt \frac{dN_\nu\{(1+z)q\}}{dq}(1+z)dq(1+z)^{-3} \tag{3}$$

$$= n_G(t_0) R_{SN}(t) dt \frac{dN_\nu\{(1+z)q\}}{dq}(1+z)dq , \tag{4}$$

where $n_G(t)$ is the number density of galaxies and $z$ is the redshift parameter, the factor $(1+z)^{-3}$ comes from the expansion of the universe, and the relation $n_G(t) = n_G(t_0)(1+z)^3$ is used. Then the total differential number density is given by the integration of Eq.(4),

$$\frac{dn_\nu(q)}{dq} = \int_{t_{min}}^{t_0} n_G(t_0) R_{SN}(t) \frac{dN_\nu\{(1+z)q\}}{dq}(1+z) dt \tag{5}$$

$$= \int_{z_{max}}^{0} n_G(t_0) R_{SN}(t) \frac{dN_\nu\{(1+z)q\}}{dq}(1+z) \frac{dt}{dz} dz . \tag{6}$$

When the cosmological constant is included, the well-known Freedman equation becomes

$$\frac{dz}{dt} = -H_0(1+z)\sqrt{(1+\Omega_0 z)(1+z)^2 - \lambda(2z+z^2)} , \tag{7}$$

where $\lambda = \frac{\Lambda}{3H_0^2}$. We now get the differential number flux of the neutrino background, $\frac{dF_\nu(q)}{dq}$, by using the relation $\frac{dF_\nu(q)}{dq} = c\frac{dn_\nu(q)}{dq}$ and Eq.(6) and (7):

$$\frac{dF_\nu}{dq} = \frac{c n_G(t_0)}{H_0} \int_0^{z_{max}} R_{SN}(t) \frac{dN_\nu\{(1+z)q\}}{dq} \frac{dz}{\sqrt{(1+\Omega_0 z)(1+z)^2 - \lambda(2z+z^2)}} . \tag{8}$$

### 3.2 Calculation of the Background Flux

According to the formula (8), we calculated the background fluxes. In the present work, we set $H_0$ to be 50 km/s/Mpc. From Eq.(8), it can be seen that the flux is proportional to $H_0^2$, because $n_G(t_0) \propto H_0^3$. Following the inflationary cosmology, we calculate for the flat universe model, $\Omega_0 + \lambda = 1$. The number density of galaxy, $n_G(t_0)$, has been estimated by some authors [22–25], but still there remains ambiguity. Here we assume that $n_G(t_0) = 0.02 h_{50}^3 [\text{Mpc}^{-3}]$, following there results. The other parameters, $z_{max}, R_{SN}(t), \frac{dN_\nu}{dq}$ are given in section 2.

In Fig.2 and Fig.3, the fluxes of the constant model and the constant+burst model are shown, respectively. In both figures, the solid line shows the flux for $\lambda = 0$, and the short and long dashed lines for $\lambda = 0.5$, $\lambda = 0.9$, as indicated. The dot-short-dashed and dot-long-dashed lines in Fig.3 show the constant component and the burst component in $\lambda = 0$ model, respectively.

In all models, the peak of flux appears at $3 \sim 4$ MeV, which is much lower than the previous work ($6 \sim 10$ MeV). It is known that the cosmological redshift effect reduces mean neutrino energy by a factor of $\frac{3}{5}$ when $\Omega_0 = 1$ and $z_{max} = \infty$ [1,2], however, the location of the peak is reduced more (Note that energy at the peak does not always give the mean energy).

In the burst model, the burst of supernovae in past shifts the energy of neutrinos to lower range, as expected. From the shape of the spectrum, we can get information about time evolution of supernova explosions, which give a constraint on the models of galaxy evolution. However, for this purpose we have to observe the low energy range where observation is difficult because of detection efficiency and reactor antineutrinos (see next subsection).

The cosmological constant raises the flux in the low energy side ($< 10$ MeV) and reduces in the high energy side ($10$ MeV $<$). The change in the spectrum is very small because the integrated number of supernovae is fixed.

## 3.3 Comparison with Other Neutrino Sources

Here we compare our results with spectra of other neutrino sources. The most serious problem in the observation of the background neutrinos from past supernovae is to exclude other backgrounds.

Fig.4 shows our flux of the relic neutrino background from past supernovae ($\lambda = 0$, constant SN rate), that of solar neutrinos [26], and that of atmospheric neutrinos [27]. We also show the flux of reactor $\bar{\nu}_e$'s at Kamioka [28] in this figure.

Note that the solar neutrinos are $\nu_e$'s, not $\bar{\nu}_e$'s, so the cross section for solar neutrinos is about two order smaller than that for $\bar{\nu}_e$'s (see next section). Furthermore, direction of recoil electrons scattered by solar neutrinos is strongly concentrated to the opposite direction of the Sun. Therefore, the solar neutrino is an avoidable background [4].

Hence, the range $10 \sim 50$ MeV is an observable range for the relic neutrinos, although lower side and upper side will be hidden by reactor neutrinos and atmospheric neutrinos, respectively.

# 4 Event Rate at the Superkamiokande Detector

## 4.1 Event Rate Calculation

The Superkamiokande (SK) and Kamiokande II (KII) are normal water Čerenkov detectors whose fiducial masses are 22000, 680 ton, respectively. They detect the characteristic transient cone of Čerenkov light from relativistic electron or positron secondaries by a dense array of inward-looking phototubes. The time integrated Čerenkov light gives the energy of recoil particles and the pattern of illuminated phototubes gives the direction of particles [4,29].

The relevant interactions of supernova neutrinos with light water are:

$$\bar{\nu}_e + p \rightarrow n + e^+ \quad \text{(C.C.)} \quad (E_{TH} = 1.8 \text{MeV}), \tag{9}$$
$$\nu_e + {}^{16}O \rightarrow {}^{16}F + e^- \quad \text{(C.C.)} \quad (E_{TH} = 15.4 \text{MeV}), \tag{10}$$
$$\bar{\nu}_e + {}^{16}O \rightarrow {}^{16}N + e^+ \quad \text{(C.C.)} \quad (E_{TH} = 11.4 \text{MeV}), \tag{11}$$
$$\nu_e + e^- \rightarrow \nu_e + e^- \quad \text{(C.C. + N.C.)}, \tag{12}$$
$$\bar{\nu}_e + e^- \rightarrow \bar{\nu}_e + e^- \quad \text{(C.C. + N.C.)}, \tag{13}$$
$$\nu_i + e^- \rightarrow \nu_i + e^- \quad \text{(N.C.)}, \tag{14}$$
$$\bar{\nu}_i + e^- \rightarrow \bar{\nu}_i + e^- \quad \text{(N.C.)}, \tag{15}$$

where (C.C.) and (N.C.) stand for the charged current and the neutral current, respectively. The subscript 'i' denotes $\mu$ or $\tau$. For cross sections of these reactions, we referred to [30–32].

When we calculate the event rate, we have to consider the detection efficiency of the detector. The expected efficiency of SK at present stage is: 0% below $T_e = 2.5$ MeV, about 50% at $T_e = 4.5$ MeV, and almost 100% above $T_e = 7.5$ MeV [33], where $T_e$ is the kinetic energy of the recoil electron (or positron). We used an appropriate fitting function in the practical calculation.

By using the flux of $\bar{\nu}_e$ obtained in the preceding section, we calculated the event rate [yr$^{-1}$MeV$^{-1}$] expected at SK for reactions (9),(11), (13) as a function of recoil electron kinetic energy, for the case of $\lambda = 0$ and constant supernova rate. We also calculate the event rate for other types of a neutrino assuming that total energy of emitted neutrinos from a supernova is the same for all six flavors and ratio of average energy (i.e. temperature) is 11 ($\nu_e$) : 16 ($\bar{\nu}_e$) : 19 (others). The ratio is based upon the results optained by Mayle et al. [34] for $25M_\odot$ model (we assume that the ratio does not depend on the stellar mass). The result is shown in Fig.5. From this result, we can neglect the contributions of the reactions other than the reaction(9). Therefore, we consider only $\bar{\nu}_e$'s and the reaction (9) in the following calculations. The event rate at KII is estimated by multiplying the volume ratio of 680/22000, since the detection efficiency properties of the two are almost the same.

The event rates are shown in Fig.6 for the constant model, and in Fig.7 for the constant+burst model. The event rate is decreased by $\Lambda$ in the observable energy range and the difference of the two models is not as conspicuous as that in flux. The rate for the constant+burst model is a little smaller than that of the constant model, because neutrinos are shifted to lower energy range, where the cross section and detection efficiency are small.

The integrated event rate at the observable range (10 $\sim$ 50 MeV) for the constant model with $\lambda = 0$ is:

$$25 \left(\frac{R_{SN}}{0.1\mathrm{yr}^{-1}}\right) \left(\frac{n_G(t_0)h_{50}^{-3}}{0.02\mathrm{Mpc}^{-3}}\right) \left(\frac{H_0}{50\mathrm{km/s/Mpc}}\right)^2 \text{ events [yr}^{-1}\text{]}. \qquad (16)$$

This formula holds essentially even in the constant+burst model, since the contribution from the burst is negligible in the range considered here.

## 4.2 Relic Neutrinos at SNO

The unique character of the Sudbury Neutrino Observatory (SNO) detector is the use of heavy water. SNO consists of 1 kton of heavy water surrounded by 1.6 kton of light water. A merit of using heavy water is that the neutral-current breakup reactions of deuterium, $\nu_i(\bar{\nu}_i) + d \rightarrow n + p + \nu_i(\bar{\nu}_i)$  (E$_{\mathrm{TH}}$ = 2.22MeV), can be used for the detection of all species of neutrinos [35].

However, the information of neutrino energy is lost because neutrinos are detected through $\gamma$ rays from absorption of the secondary neutrons in $^{37}$Cl atoms. As shown in Fig.4, it is difficult to distinguish the relic neutrinos from other backgrounds if there is no information about the energy. Therefore, the neutral current reactions is not useful for the detection of the relic neutrinos from supernovae.

Of course, there are also the charged current reactions as in the light water detector, by which the neutrino energy can be measured. The cross sections of these reactions are, however, a little smaller than that of the reaction (9) [36]. Considering the mass of water, the event rate at SNO detector is at most the same order with that of KII detector. In conclusion, SNO is not suitable for observation of the relic neutrino background.

## 5 Discussion and Conclusions

We calculated the spectrum and the event rate at the Superkamiokande detector more systematically than any previous work. The spectrum has its peak at neutrino energy 3 $\sim$ 4 MeV, which is much lower than previous work (6 $\sim$ 10 MeV). The event rate at SK in the observable range (10 $\sim$ 50 MeV) is about $25\, h_{50}{}^2 \left(\frac{R_{SN}}{0.1\mathrm{yr}^{-1}}\right) \left(\frac{n_G h_{50}^{-3}}{0.02\mathrm{Mpc}^{-3}}\right)$ events per year, where $\Omega_0 = 1$, $\lambda = 0$, and the supernova rate is constant with time. The amount of the flux depends upon mainly the Hubble constant and the supernova rate, so if the amount is determined from observations, we get an estimate of $H_0^{-1} R_{SN} n_G$.

We compared two models about time dependence of the supernova rate. It is found that bursts of supernovae in the formation epoch of galaxies can change the shape of the spectrum especially in the low energy range ($\lesssim$ 10MeV), however, does not strongly affect in high energy range, provided the recent supernova rate is constant. It is quite important to know time dependence of the supernova rate because it strongly reflects the history of galaxies. We can infer the time dependence from the shape of the spectrum, however, for this purpose we have to observe the low energy range where observation is more difficult because of detection efficiency and reactor $\bar{\nu}_e$'s.

The effect of a cosmological constant is also investigated. The spectrum is almost insensitive to a cosmological constant, if the time integrated number of supernovae in a galaxy is fixed, and more sensitive to the Hubble constant ($\propto H_0^2$). Therefore, it is almost impossible to use the flux as a test for existence of $\Lambda$.

Since SK has good energy resolution, we will be able to measure the average effective temperature of supernova neutrinos from the slope of the spectrum, which is free from uncertainty of the Hubble constant, the supernova rate, and the number density of galaxies. This temperature measurement will provide us valuable information to the stellar mass function (temperature depends upon the mass of progenitors), the model of supernova explosions, etc.

The largest uncertainty in the present work is, of course, the supernova rate. In order to get a reasonable model of the supernova rate, we must consider a reliable model of evolution of galaxies. Such calculations will be given in our next paper, in which the supernova rate is given by a reasonable model of galaxy evolution based upon the population synthesis method, which agrees with observations well.

# Acknowledgement

We would like to thank Y. Totsuka for valuable discussion and for providing us with the useful data on the Kamiokande detectors.

Table 1: Supernova Rates for Various Models

| | cosmic time [Gyr] ($z = 5$ to $z = 0$) | SN rate[a] of constant component (constant) | (constant+burst) | SN rate[a] at burst peak (constant+burst) |
|---|---|---|---|---|
| $\lambda = 0.0$ | 12.1 | $8.26 \times 10^{-2}$ | $4.13 \times 10^{-2}$ | $7.55 \times 10^{-1}$ |
| $\lambda = 0.5$ | 15.0 | $6.67 \times 10^{-2}$ | $3.34 \times 10^{-2}$ | $5.42 \times 10^{-1}$ |
| $\lambda = 0.9$ | 22.2 | $4.50 \times 10^{-2}$ | $2.25 \times 10^{-2}$ | $2.70 \times 10^{-1}$ |

[a] unit yr$^{-1}$ per a galaxy
$H_0 = 50$ km/s/Mpc, $\Omega_0 + \lambda = 1$.
Total number of supernovae is $10^9$ per a galaxy.

# Figure Captions

**Figure 1:** The weighted average of the differential number of neutrinos emitted from a supernova (solid line). The spectra for $10M_\odot, 15M_\odot, 25M_\odot$ are also shown [3].

**Figure 2:** The relic neutrino number fluxes of the constant model with some values of $\lambda$ are shown. It is assumed that the universe is flat, i.e. $\Omega_0 + \lambda = 1$.

**Figure 3:** The relic neutrino number fluxes of the constant+burst model with some values of $\lambda$ are shown. The constant and burst components when $\lambda = 0$ are also shown. As in Fig.2, it is assumed that $\Omega_0 + \lambda = 1$.

**Figure 4:** The solid line shows the relic neutrino flux from past supernovae (the flux when $\lambda = 0$, shown in Fig.2). The two dashed lines show the fluxes of the $^8$B and hep solar neutrinos, as indicated [26]. The dot-short-dashed line shows the flux of the atmospheric neutrinos ($\bar{\nu}_e$, at Kamiokande) [27], and the dot-long-dashed line shows the flux of reactor $\bar{\nu}_e$'s [28].

**Figure 5:** The event rate at the Superkamiokande detector for relevant reactions. The used flux is that of constant model, $\lambda = 0.0$. Temperature difference among neutrino flavors are taken into consideration. The three solid lines show the event rate for the reactions $\bar{\nu}_e p \to n e^+$, $\bar{\nu}_e\,^{16}O \to\,^{16}N\,e^+$, $\nu_e\,^{16}O \to\,^{16}F\,e^-$, as indicated. The short-dashed line gives the event rate for the reaction $\nu_e e^- \to \nu_e e^-$, the long-dashed line for $\bar{\nu}_e$, the dot-short-dashed line for $\nu_\mu$ (or $\nu_\tau$), the dot-long-dashed line for $\bar{\nu}_\mu$ (or $\bar{\nu}_\tau$).

**Figure 6:** The event rates of the constant model at SK, corresponding fluxes are shown in Fig. 2.

**Figure 7:** The event rates of the constant+burst model at SK, corresponding fluxes are shown in Fig. 3.

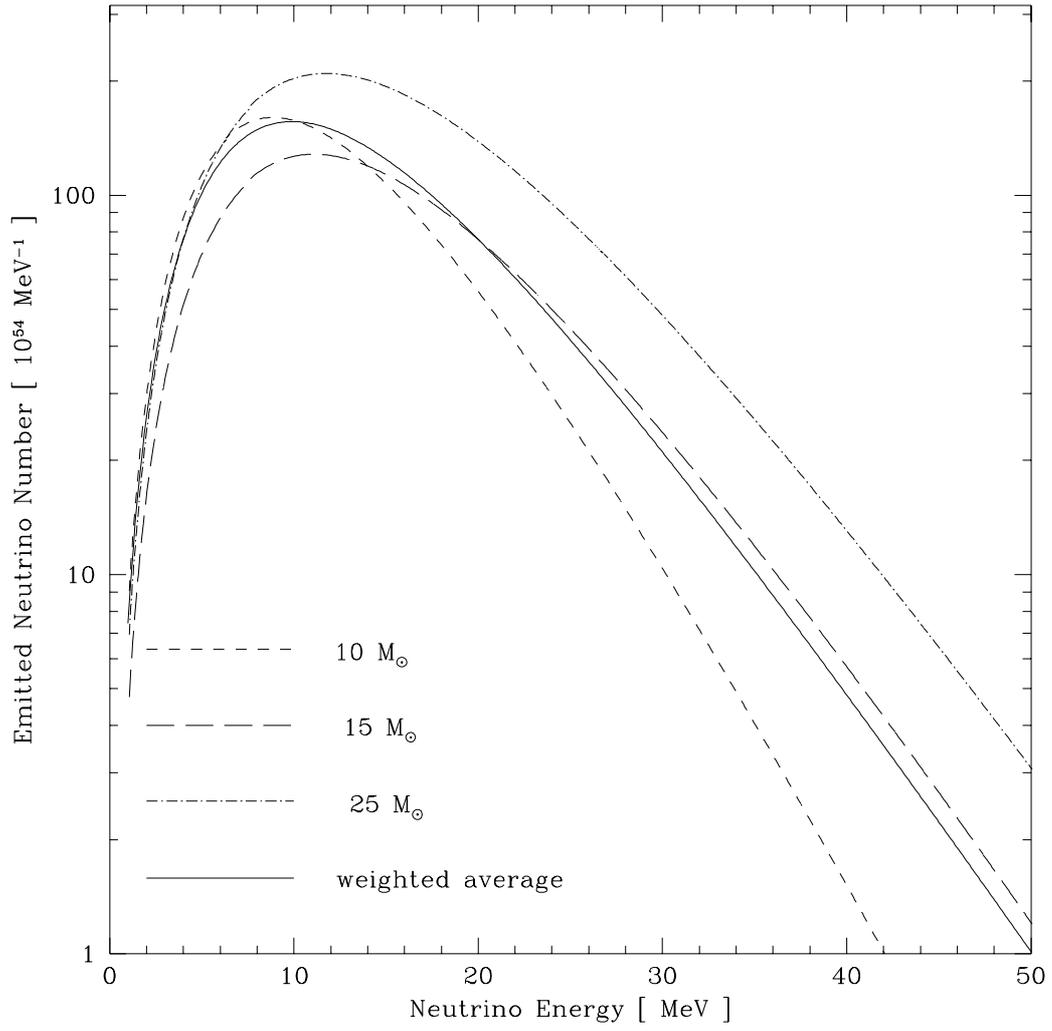

Figure 1:

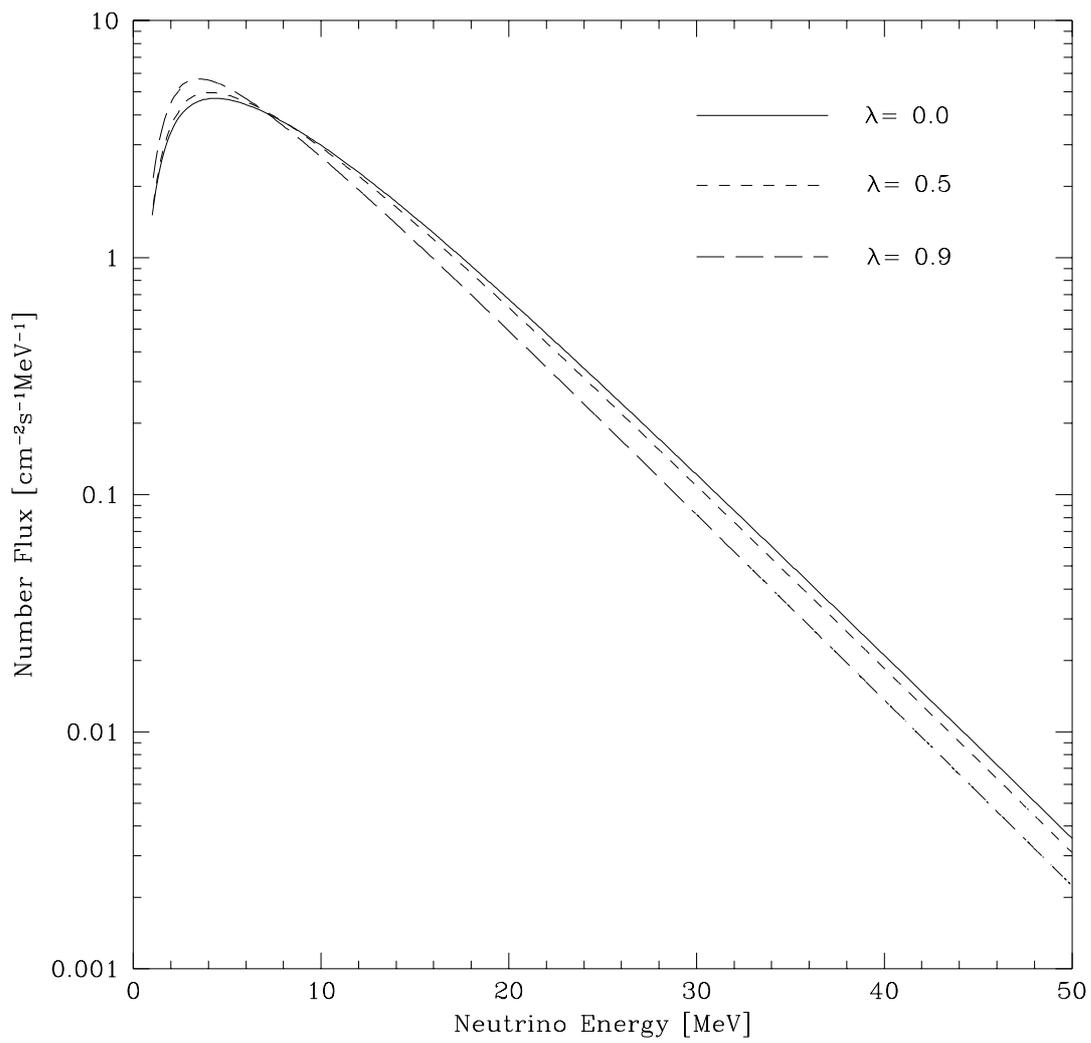

Figure 2:

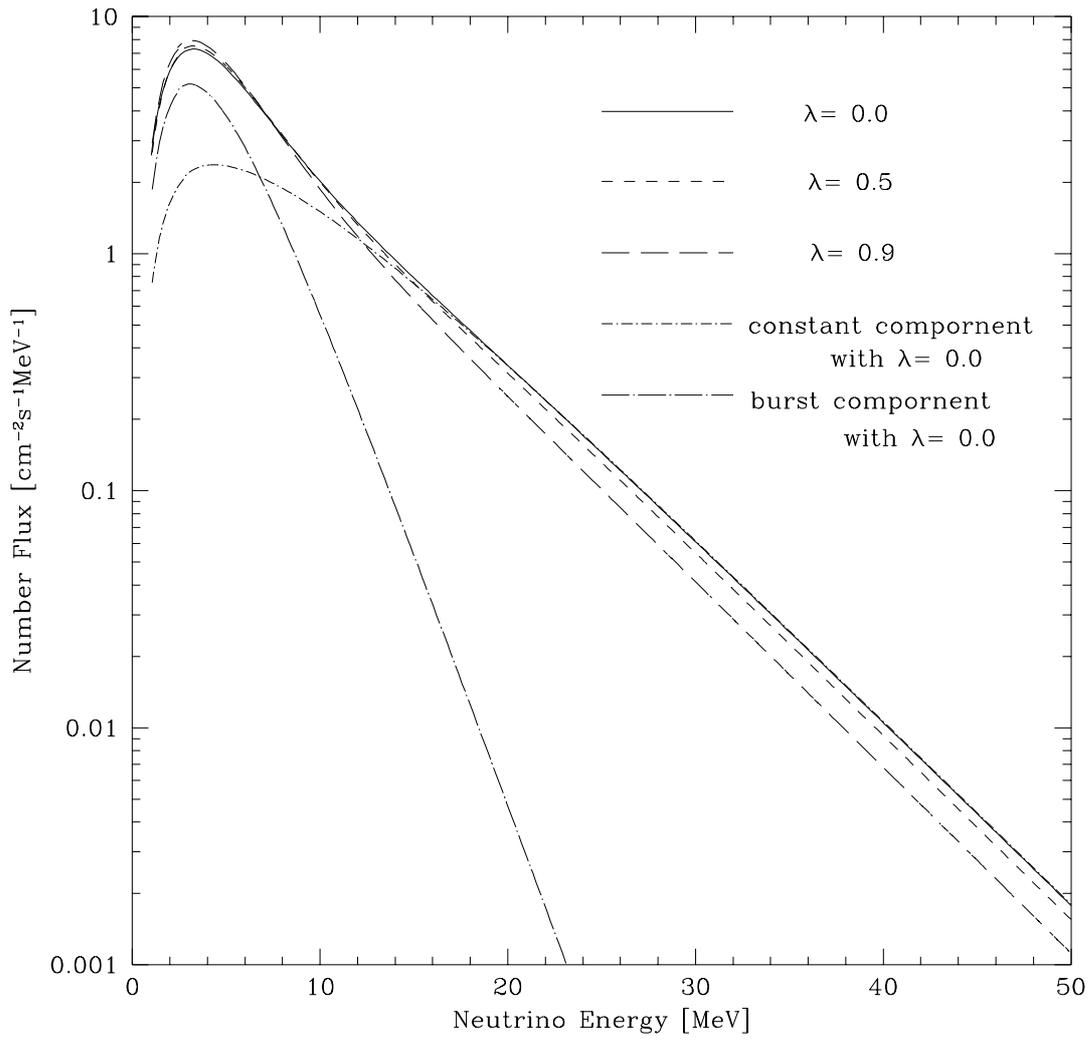

Figure 3:

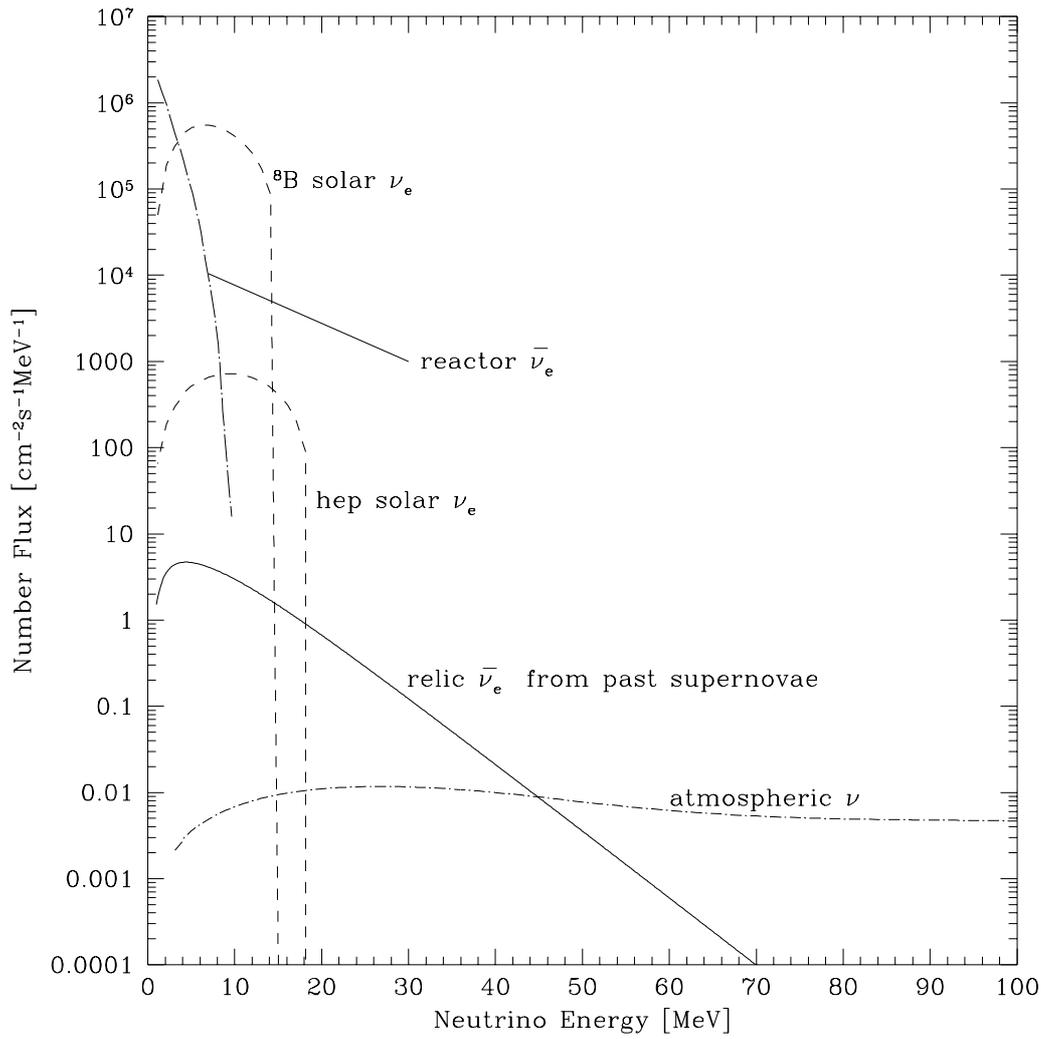

Figure 4:

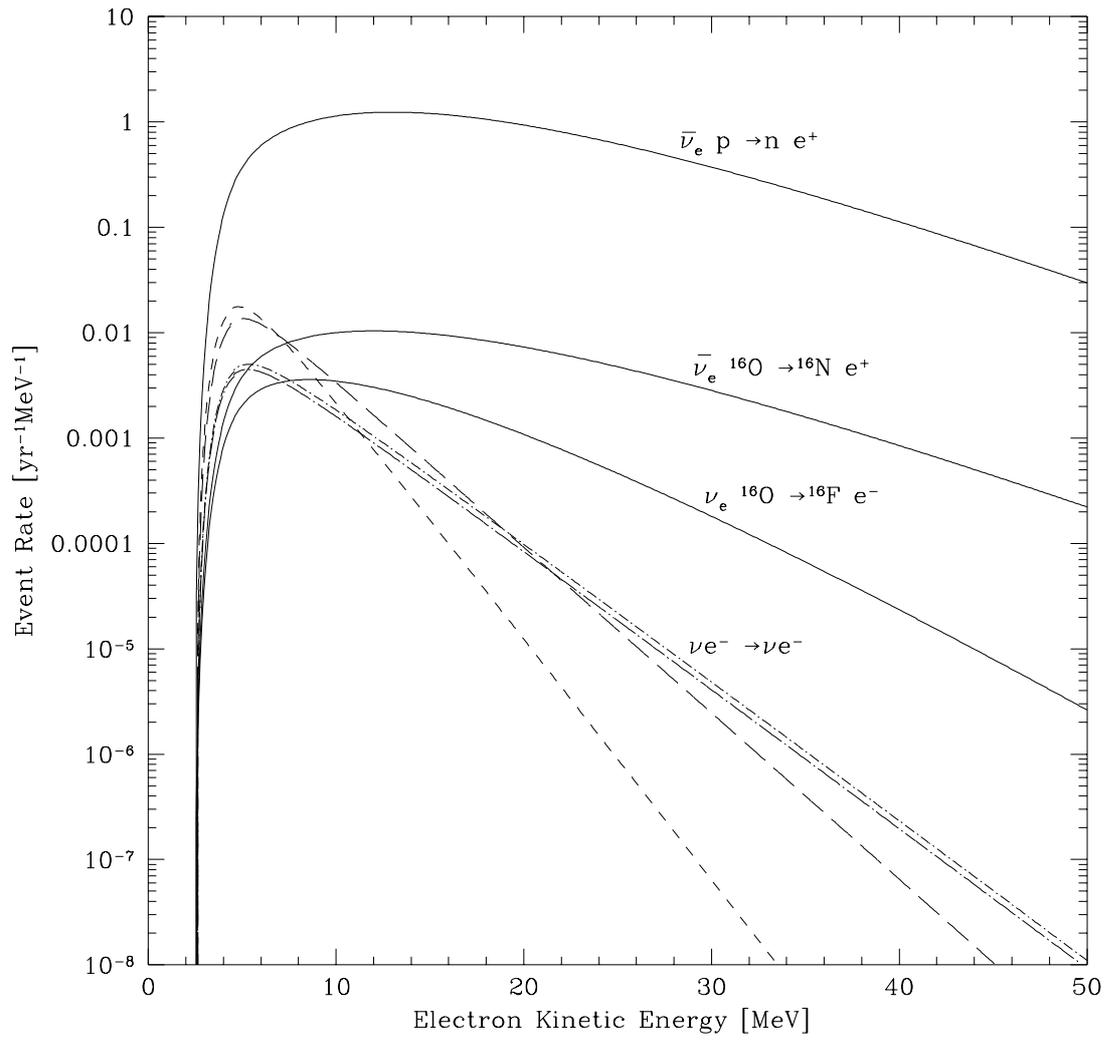

Figure 5:

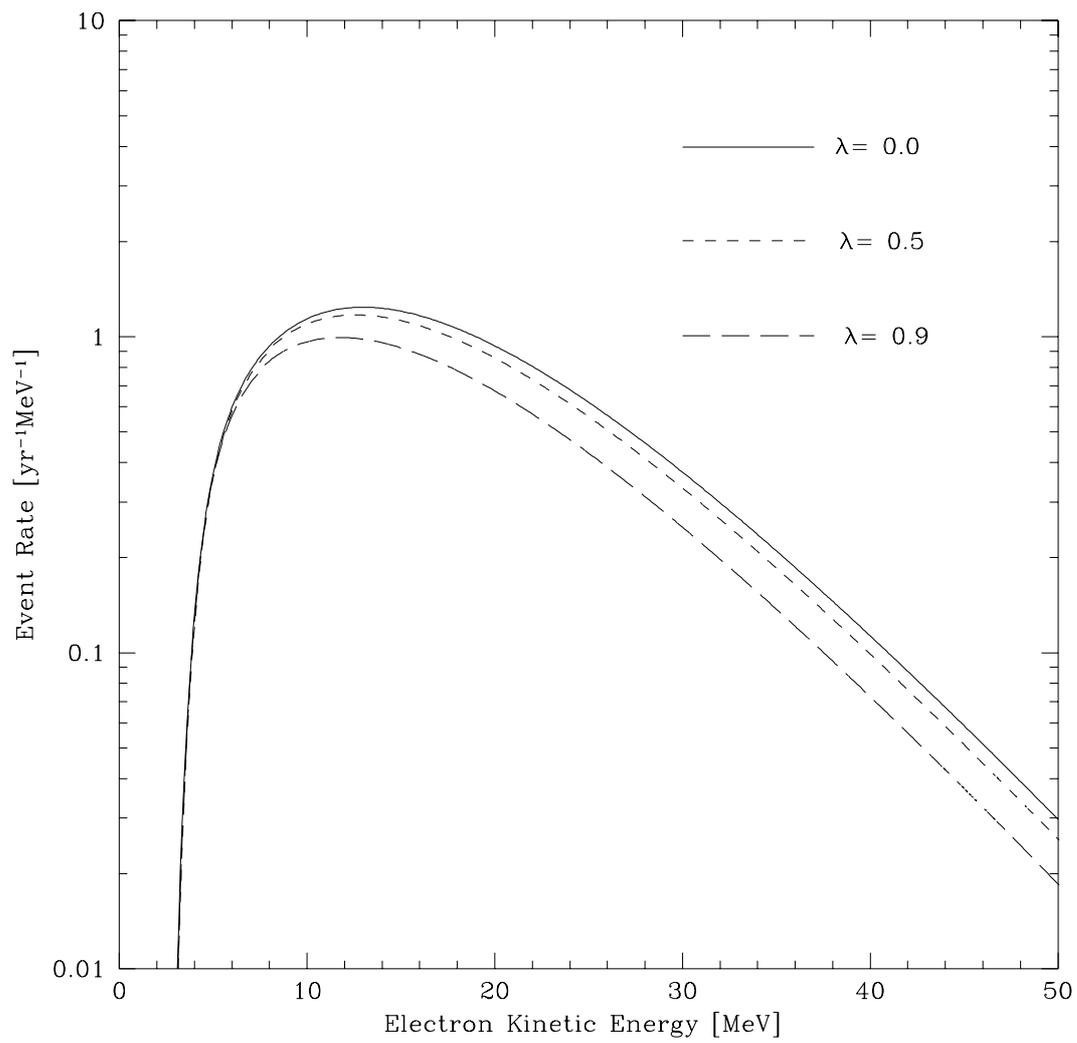

Figure 6:

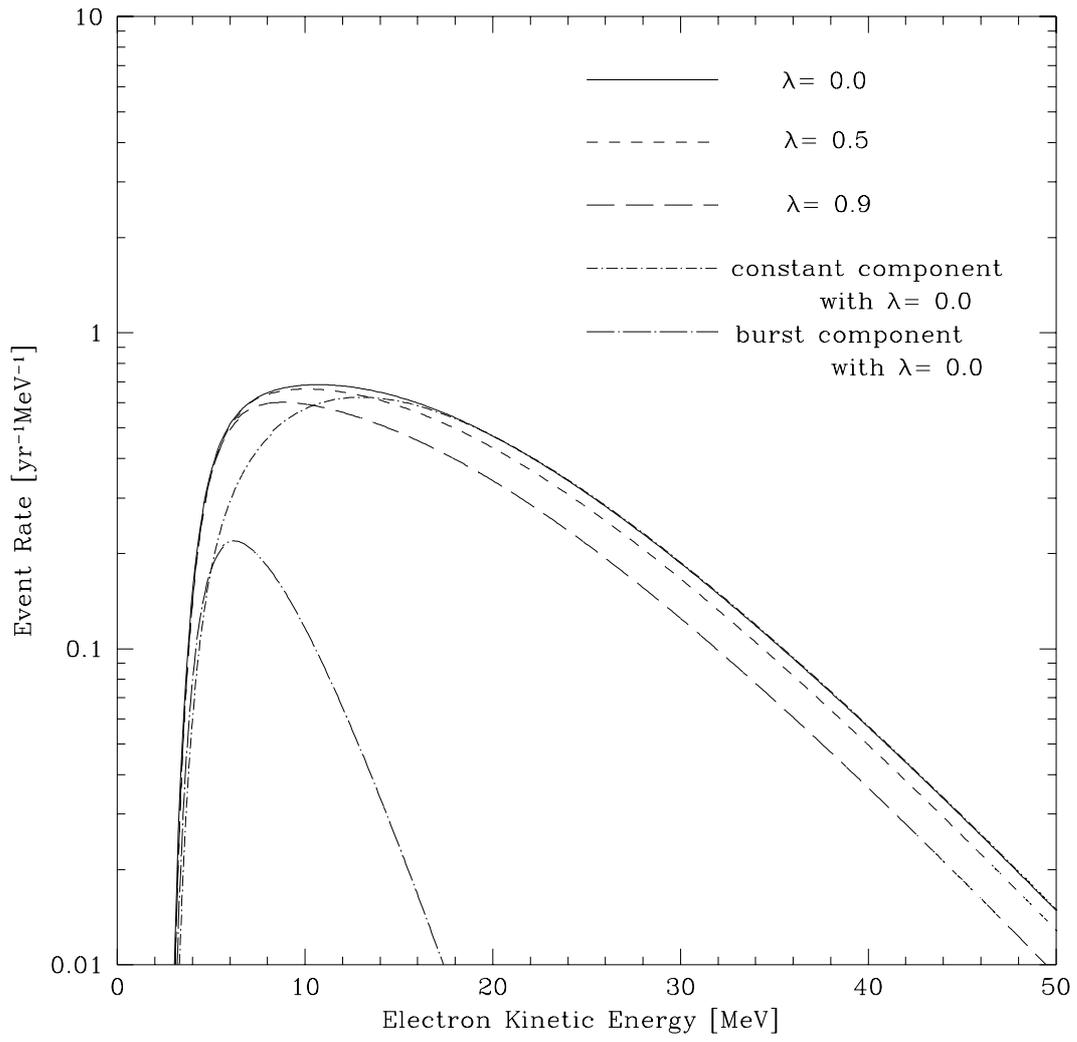

Figure 7: